\documentclass[conference]{IEEEtran}
\usepackage{balance}

\usepackage{cite}

\ifCLASSINFOpdf
  \usepackage[pdftex]{graphicx}
  
\else
  \usepackage[dvips]{graphicx}
\fi

\usepackage[cmex10]{amsmath}
\usepackage{array}
\usepackage{mdwmath}
\usepackage{mdwtab}
\usepackage{amssymb}
\usepackage{bm}
\usepackage{mathrsfs}
\usepackage{filecontents}
\usepackage{stfloats}
\usepackage{color}
\definecolor{ao}{rgb}{0.0, 0.5, 0.0}
\definecolor{copper}{rgb}{0.72, 0.45, 0.2}
\usepackage{footmisc}
\ifCLASSOPTIONcompsoc
	\usepackage[caption=false,font=normalsize,labelfont=sf,textfont=sf]{subfig}
\else
	 \usepackage[caption=false,font=footnotesize]{subfig}
\fi

\usepackage{stfloats}
\usepackage{url}
\newtheorem{theorem}{Theorem}
\newtheorem{corollary}{Corollary}
\newtheorem{Lemma}{Lemma}

\newcommand{\goodchi}{\protect\raisebox{2pt}{$\chi$}}

\newtheorem{definition}{Definition}

\newtheorem{proposition}{Proposition}

\begin{document}
\IEEEoverridecommandlockouts
\title{Graph-Theoretic Approaches to Two-Sender Index Coding}
\author{\IEEEauthorblockN{Chandra Thapa, Lawrence Ong, and Sarah J. Johnson}
	\IEEEauthorblockA{School of Electrical Engineering and Computer Science, The University of Newcastle, Newcastle, Australia\\
		Email: chandra.thapa@uon.edu.au, lawrence.ong@cantab.net, sarah.johnson@newcastle.edu.au}
	\thanks{This work is supported by the Australian Research Council under
		grants FT110100195, FT140100219, and DP150100903.} 
	}
\maketitle

\begin{abstract}
 Consider a communication scenario over a noiseless channel where a sender is required to broadcast messages to multiple receivers, each having side information about some messages. In this scenario, the sender can leverage the receivers' side information during the encoding of messages in order to reduce the required transmissions. This type of encoding is called index coding. In this paper, we study index coding with two cooperative senders, each with some subset of messages, and multiple receivers, each requesting one unique message. The index coding in this setup is called two-sender unicast index coding (TSUIC). The main aim of TSUIC is to minimize the total number of transmissions required by the two senders. Based on graph-theoretic approaches, we prove that TSUIC is equivalent to single-sender unicast index coding (SSUIC) for some special cases. Moreover, we extend the existing schemes for SSUIC, viz., the cycle-cover scheme, the clique-cover scheme, and the local-chromatic scheme to the corresponding schemes for TSUIC.

\end{abstract}


\IEEEpeerreviewmaketitle

\section{Introduction}

In a noiseless broadcast-channel setting, a sender can utilize the knowledge of messages available at each receiver a priori, and broadcasts coded symbols (e.g., bit-wise XOR of two message symbols) to reduce the number of transmissions required. The messages available at each receiver a priori is called its side information. Encoding by exploiting side information at receivers is known as index coding, and the resulting code is called an index code. Finding the optimal, i.e., the shortest, index code for any configuration of side information is called the index-coding problem. 
The index-coding problem was introduced by Birk and Kol~\cite{ISCOD}, and further studied in subsequent works~\cite{maisbound,broadcastrate,localgarphcoloring,chaudhary,neely,interferencealignment,composite,structuralproperties,linearcodeoptimal,ourpaper2}.

Most of the earlier works on index coding considered only one sender. However, there exist communication problems where a set of messages are distributed among multiple senders. For example, if multiple ground stations are in the coverage of two satellites each having a subset of messages (possibly with some overlapping messages) required by the ground stations. The index-coding problem with multiple senders is called the multi-sender index-coding problem. It is further motivated by its relation to the \emph{cooperative-data-exchange} problem, which is defined as follows: For a message set $ \mathcal{M}$, each receiver $ r\in \{1,2,\cdots,N\} $ possesses a subset of messages $ \mathcal{H}_r\hskip-1.5pt\subset\hskip-1.5pt \mathcal{M}$, and wants to receive all other messages in $ \mathcal{M}\setminus \mathcal{H}_r$ by communicating with other receivers. All receivers cooperate with each other and communicate by transmitting coded symbols such that all demands made by all the receivers are fulfilled with the minimum number of total transmissions. The problem of finding the minimum number of total transmission in this setup is called the cooperative-data-exchange problem~\cite{coopdata1}. This problem can be viewed as a special case of the multi-sender index-coding problem in the following way: There are $ N$ senders and $ N $ receivers, each receiver $ r\in \{1,2,\cdots,N\}$ has a subset of messages $ \mathcal{H}_r$ and request messages $ \mathcal{M}\setminus \mathcal{H}_r  $ from $ N $ senders, where each sender $ S_i $, for $ i\in \{1,2,\cdots,N\}  $, has the message set $\mathcal{H}_i$. The solution of this index-coding problem is the solution of the corresponding cooperative-data-exchange problem, and vice versa. 

Multi-sender index coding was first studied by Ong et al.~\cite{lawrencemultisender}. Their system considered multiple senders, multiple receivers, and each receiver knows only one message a priori. Furthermore, the senders can collaborate in order to transmit a shortest index code. The index coding in this setup is called the multi-sender single-uniprior index-coding. Based on graph-theoretic approaches, they gave a lower bound and an upper bound, which is based on cyclic codes, on the shortest index codelength. They proved that the bounds match for some special cases, including the case where no two senders know any message in common. In a subsequent work by Sadeghi et al.~\cite{parastoomultisender}, multi-sender index coding was further studied in a different setup. Their system considered $ 2^N-1 $ senders, $ N $ receivers, and each receiver, knowing some subset of messages, wants one unique message. The index coding in this setup is called distributed index-coding. Based on a random-coding approach, they proposed a distributed composite-coding scheme, which was built upon the existing single-sender composite-coding scheme~\cite{composite}. They solved all instances of the problem with up to $ N=3 $.  

In this paper, we consider a similar setup to the multi-sender single-uniprior index-coding setup~\cite{lawrencemultisender} except that in our system, we consider two senders and each receiver has some subset of messages (possibly more than one message) a priori. We call index coding on this setup two-sender unicast index coding (TSUIC). The problem of finding the minimum number of total transmissions required by the two senders is called the TSUIC problem. This work is a step towards understanding multi-sender index-coding problems. To the best of our knowledge, TSUIC has not been studied before. 

\subsection{Our contributions}

We prove that for some special cases, TSUIC is equivalent to single-sender unicast index coding (SSUIC). We next extend
the cycle-cover scheme to TSUIC problems. We prove that only certain cycles, which we call \emph{message-connected} cycles, can be exploited in TSUIC to obtain some savings.\footnote{\label{f4}The number of transmissions saved by transmitting coded symbols rather than transmitting uncoded message is called the \emph{savings} of the index code.} For a special case, we show that the two-sender cycle-cover scheme achieves the single-sender cycle-cover codelength. We then extend the clique-cover scheme to TSUIC problems. We prove that the two-sender clique-cover scheme achieves the codelength equal to the chromatic number of one special graph constructed from the union of a graph representing the senders' message setting (which we call the sender-constraint graph) and the complemented digraph\footnote{\label{f3}The complemented digraph of a digraph $ D $, represented as $ \bar{D} $, is a digraph with (i) $ V(\bar{D})=V(D) $ and (ii) for any two vertices $i,j\in V(\bar{D}) $, $ (i,j)\in A(\bar{D}) $ if and only if $ (i,j)\notin A(D) $.} of a digraph representing the receivers' side information. Furthermore, for a special case, we show that the two-sender clique-cover scheme achieves the single-sender clique-cover codelength. Finally, we extend the local-chromatic-number scheme to TSUIC problems. We first extend the concept of local-chromatic number to suit the two-sender setting similar to the way we extend the cycle-cover and the clique-cover schemes. However, unlike those cases, the two-sender local-chromatic number is not always achievable. So, in TSUIC, we provide an achievable index codelength that is a function of both the two-sender local-chromatic number and colors on the sender-constraint graph. The bound is further tightened by applying the partitioned approach.	
	
\section{Problem setup and Definitions}
We consider a TSUIC problem with the following setup:
\begin{itemize}
	\item $ N $ independent messages, represented by $ \mathcal{M}=\{x_1,x_2,\dotsc,x_N\} $, and each message $ x_i\in \{0,1\}^t $ for all $ i $ and some integer $ t\geq 1 $.
	\item Two senders, denoted $ S_1 $ and $ S_2 $, have (ordered\footnote{\label{f1}The elements are ordered in increasing indices.}) message sets $\mathcal{M}_1 \subseteq \mathcal{M}$ and $\mathcal{M}_2 \subseteq \mathcal{M}$ respectively such that $\mathcal{M}_1\cup \mathcal{M}_2=\mathcal{M} $ (i.e., each message is available at some sender(s)).  
	\item $ N $ receivers,  where each receiver $r\in \{ 1,2,\dotsc,N\}$ has a subset of messages a priori, represented by an ordered\footref{f1} set $ \mathcal{H}_r \subseteq \mathcal{M}\setminus \{x_r\} $, and wants a distinct message $ x_r $. The messages in $\mathcal{H}_r $ are side information at $ r $. 
\end{itemize}
This setup is called the TSUIC setup because we have two senders and multiple receivers each requesting one unique message from the senders. Now we define a two-sender index code for the above setup: 
\begin{definition} [Two-sender index code]
	A two-sender index code is a set ($ \{\mathscr{F}_s\},\{\mathscr{G}_r\} $) which consists of
	\begin{enumerate}
		\item an encoding function for each sender $ S_s $, $ \mathscr{F}_{s}: \{0,1\}^{|\mathcal{M}_{s}|\times t}\rightarrow \{0,1\}^{\ell_{s}}$ such that $\mathcal{C}_s= \mathscr{F}_{s}(\mathcal{M}_{s})$, and   
		\item a decoding function for every receiver $ r $, $ \mathscr{G}_r: \{0,1\}^{(\Sigma_{s=1}^2 \ell_{s} + |\mathcal{H}_r| \times t)}\rightarrow \{0,1\}^t $ such that $ x_r = \mathscr{G}_r(\mathcal{C}_1,\mathcal{C}_2,\mathcal{H}_r) $.  
	\end{enumerate}
\end{definition}
This means each sender $ S_s $ encodes its messages to a total of $ \ell_{s}$-bits coded symbols for some integer $ \ell_s $. 
We assume that each receiver $ r $ receives all $\mathcal{C}_s $ without any noise, and it decodes $x_r $ from the received coded symbols and $ \mathcal{H}_r $.

\begin{definition}[Index codelength]
	The index codelength of a two-sender index code ($ \{\mathscr{F}_s\},\{\mathscr{G}_r\} $) is the total number of transmitted bits per received message bits, i.e., $ \ell=\frac{(\ell_{1}+\ell_{2})}{t}$. This is also referred as the (normalized) length of the index code. We say that a $ \ell $ is achievable for a TSUIC problem if there exists an index code of normalized length $ \ell $.  
\end{definition}

\begin{definition} [Optimal broadcast rate]
	The optimal broadcast rate of a TSUIC problem with $ t $-bit messages is $ \beta_t\triangleq  \underset{\{\mathscr{F}_{s}\},\{\mathscr{G}_r\}}{\mathrm{inf}}\ \ell$, and the optimal broadcast rate over all $ t $ is defined as $ \beta=\underset{t}{\mathrm{inf}}\ \beta_{t}$.  
\end{definition}

In TSUIC, $ S_1 $ (sender one) encodes the messages in $ \mathcal{M}_1 $, and $ S_2 $ (sender two) encodes the messages in $ \mathcal{M}_2 $ in order to achieve the broadcast rate $ \beta_{t} $. In general, each sender has \emph{private messages}\footnote{The private messages are messages present only at one sender.} and \emph{common messages}.\footnote{The common messages are messages present at both senders.} Let $ \mathcal{P}_1 =\mathcal{M}_1\setminus\mathcal{M}_2$ and $ \mathcal{P}_2 =\mathcal{M}_2\setminus\mathcal{M}_1$ be the set of private messages at senders $ S_1 $ and $ S_2 $ respectively, and $ \mathcal{P}_c =\mathcal{M}_1\cap \mathcal{M}_2$ be the set of common messages at both senders. We now define one constraint in TSUIC.
\begin{definition} [Constraint due to the two senders]
	The constraint due to the two senders is the following: While encoding, any two private messages $ x_i\in \mathcal{P}_1 $ and $ x_j\in \mathcal{P}_2 $ should not be encoded together (with or without other messages) to construct one coded symbol, or alternatively $\mathcal{C}_1= \mathscr{F}_{1}(\mathcal{M}\setminus \mathcal{P}_2 )$ and $\mathcal{C}_2= \mathscr{F}_{2}(\mathcal{M}\setminus \mathcal{P}_1)$. 
\end{definition}
\begin{figure}[t] 
	\centering
	\subfloat []{ %
		\includegraphics[height=1.5cm,keepaspectratio]{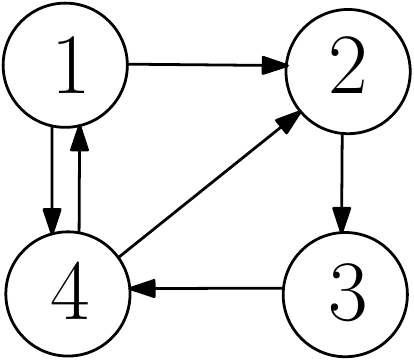}
		\label{fig1}}
	\hspace{1cm}
	\subfloat []{
		\centering
		\includegraphics[height=1.5cm,keepaspectratio]{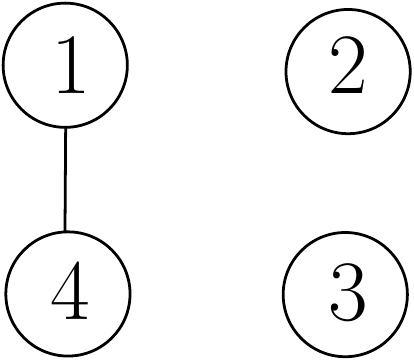}
		\label{fig2}
	}
	\caption{(a) The digraph $ D $ representing the receivers' side information of an instance of unicast index-coding problems, and (b) the sender-constraint graph $ G_o $ due to message sets $ \mathcal{M}_1 =\{ x_1,x_2,x_3\} $ and $ \mathcal{M}_2 =\{ x_2,x_3,x_4\} $ at $ S_1 $ and $ S_2 $ respectively. In $ G_o $, vertices 1 and 4 are connected by one edge as neither sender contains both $ x_1 $ and $ x_4 $. Consequently, any codeword with $ x_1\oplus x_4 $ is an invalid codeword, where $ \oplus $ indicates bit-wise XOR.}
	\label{fig12}
\end{figure}  
\subsection{Graphical representation of the receivers' side information and the senders' message setting}
  For a unicast index-coding problem, the receivers' side information can be represented by a digraph $ D=(V(D),A(D) )$, where $ V(D)=\{1,2,\dotsc,N\} $ is a set of vertices representing the $ N $ receivers, and an arc $ (i\rightarrow j) \in A(D)$ exists from vertex $ i $ to $ j $ if and only if $ x_j\in \mathcal{H}_i$. Now to reflect the senders' message setting in TSUIC, we introduce an undirected graph, denoted by $ G_o=(V(G_o),E(G_o))$, that is constructed in the following way:
(i) $ V(G_o)=V(D)$, and (ii) for $ i,j\in V(G_o) $, an undirected arc, i.e., an edge $ (i,j)\in E(G_o)$ if and only if $ x_i\in \mathcal{P}_1$ and $ x_j\in \mathcal{P}_2$, or vice versa. This means, there is an edge connecting two vertices in $ G_o $ if and only if the messages requested by those vertices are not both available at any single sender. We call the graph $ G_o $ the \emph{sender-constraint graph}. For example, refer to Fig.~\ref{fig12}. 

\begin{definition}[Maximum acyclic induced sub-digraph \textnormal{(MAIS)}]
	For a digraph $ D $, an induced acyclic sub-digraph formed by removing the minimum number of vertices in $ D $, is called a maximum acyclic induced sub-digraph (MAIS). The order of the MAIS is denoted as $ \mathsf{MAIS}(D) $. 
\end{definition}

As a TSUIC problem is described by $ D $ and $ G_o $, it is represented by $ (D,G_o) $ in this paper. Given a $ (D,G_o) $, let $ \ell(D,G_o) $ denote the index codelength, $ \beta_t(D,G_o)$ and $\beta(D,G_o)$ represent the optimal broadcast rate and the optimal broadcast rate over all $ t $ respectively. $ \ell(D) $, $ \beta_t(D)$ and $\beta(D)$ are the respective terms used for single-sender cases. We remark that, for any $ (D,G_o) $, we have $ \beta(D)\leq \beta(D,G_o) $.

\section{Results}

\begin{proposition} \label{theorem1}
	If any sender has all of the messages in $ \mathcal{M} $, then the TSUIC problem is equivalent to the SSUIC problem.
\end{proposition}

For the given condition, one can prove the Proposition~\ref{theorem1}. Now for a special case of senders' message settings, we prove that the TSUIC problem is equivalent to two separate SSUIC problems.\vskip+2pt
\begin{theorem} \label{theorem2}
	If $\mathcal{P}_c=\emptyset $, then $ \beta_t(D,G_o)= \Sigma_{s=1}^2 \beta_t(D_s)$ where $ D_s $, the sub-digraph\footnote{\label{f2}We say a digraph $ D_1=(V(D_1),A(D_1))$ is a vertex-induced sub-digraph of a digraph $ D=(V(D),A(D)) $ if $ V(D_1)\subseteq V(D) $ and all the arcs between the vertices in $ V(D_1) $ from $ A(D) $ are in $ A(D_1) $.} of $ D $, induced by $ V(D_s)=\{i: x_i\in \mathcal{M}_s \} $, for $ s\in \{1,2\} $.
\end{theorem}
\begin{IEEEproof} 
For the sub-digraphs $ D_1$ and $ D_2 $, which are induced by the vertex sets $ V(D_1)$ and $ V(D_2)$ respectively, let their sender-constraint graphs be represented by $ G'_o $ and $ G''_o $ respectively. As $ \mathcal{P}_c=\emptyset $, so $ G'_o=(V(D_1),\emptyset) $, $ G''_o=(V(D_2),\emptyset)$, and $ V(D)=V(D_1)\cup V(D_2) $. Thus we get $ \beta_t(D_1,G'_o) +\beta_t(D_2,G''_o)=\beta_t(D_1)+\beta_t(D_2)$. Now for the given TSUIC, we have $ \beta_t(D,G_o)\leq \beta_t(D_1,G'_o)+\beta_t(D_2,G''_o)=\beta_t(D_1)+\beta_t(D_2) $ because the broadcast rate for a digraph is less than or equal to the sum of the broadcast rates of its sub-digraphs covering all of the vertices of the digraph. Moreover, $ G_o $ is a complete bipartite graph  $ K_{|V(D_1)|,|V(D_2)|} $, thus on the senders' end, $\mathcal{C}_1= \mathscr{F}_{1}(\mathcal{P}_1)$ and $\mathcal{C}_2= \mathscr{F}_{2}(\mathcal{P}_2)$. Consequently, on the receivers' end, any vertex $ i\in V(D_1) $ cannot exploit $\mathcal{C}_2 $ in decoding $ x_i $ because $\mathcal{C}_2= \mathscr{F}_{2}(\mathcal{P}_2)$ and $ \mathcal{P}_2\cap \mathcal{M}_1=\emptyset $. Similarly, any vertex $ j\in V(D_2) $ cannot exploit $\mathcal{C}_1 $ in decoding $ x_j $. Thus $ \beta_t(D,G_o)=\beta_t(D_1,G'_o) +\beta_t(D_2,G''_o)$. Altogether, $ \beta_t(D,G_o)=\beta_t(D_1)+\beta_t(D_2)$.    	
\end{IEEEproof}

\subsection{A Trivial extension of SSUIC schemes to TSUIC schemes}

\begin{proposition} \label{prop2}
	Given $ D$, if $ \ell_X(D) $ is achievable using any scheme $ X $ in SSUIC, then $ \sum_{i}^{}\ell_X(D_i)$ is achievable in TSUIC by using the same scheme $ X $ on each $ D_i $, where $ D_i $ is the vertex-induced sub-digraph\footref{f2} satisfying (i) $ V(D_i)\subseteq \{ k: x_k\in \mathcal{M}_1\}$ or $V(D_i)\subseteq \{k: x_k\in \mathcal{M}_2\}$, (ii) $ V(D_i)\cap V(D_j)=\emptyset$ for $ i\neq j $, and (ii) $ \bigcup_i V(D_i)=V(D)$. 
\end{proposition}
\begin{IEEEproof}
	In SSUIC, for any scheme $ X $, $ \ell_X(D_i) $ is achievable for any given sub-digraph $ D_i $. Let $ \mathcal{A}_i $ be the set of coded symbols generated by the scheme $ X $ achieving $ \ell_X(D_i) $ for $ D_i $. As any vertex $ j \in V(D) $ has at least as much side-information as $j$ in $ D_i $ that contains $ j $, so $ \bigcup_i \mathcal{A}_i $ is an index code for $ D $, and any $ j\in V(D) $ can decode $ x_j $ from $ \bigcup_i \mathcal{A}_i $ and $ \mathcal{H}_j$. Thus, $ \sum_{i}^{}\ell_X(D_i)$ is achievable for $ D $. In TSUIC, for the given sub-digraphs of $ D $, the corresponding sender-constraint graph for each $ D_i $ is $ (V(D_i), \emptyset) $, i.e., with no edge. This means, for each $ D_i $, the set of requested messages $ \{x_k:k \in V(D_i)\} $ is either subset of $ \mathcal{M}_1 $ or $ \mathcal{M}_2 $. Thus for each $ D_i $, there is no constraint due to the two senders. Therefore, $ \bigcup_i \mathcal{A}_i $ can be transmitted on TSUIC to achieve $ \sum_{i}^{}\ell_X(D_i)$. 
\end{IEEEproof}

In SSUIC, cycles and cliques present in $ D $ are two important graph structures that provide savings\footref{f4}\cite{ISCOD,chaudhary}. Now in the following sections, we extend them to TSUIC.

\subsection{Cycle cover and its two-sender extension}

In SSUIC, the cycle-cover scheme\cite{chaudhary,neely} exploits any cycle $ C $ in $ D $ to get some savings.\footref{f4} Moreover, if $ C $ has $ n $ vertices, then for $ C $, the scheme gives a set of coded symbols
of length $ n-1 $. However, in TSUIC, because of the introduction of the constraint due to the two senders, we can only exploit a cycle $ C $ which is \emph{message connected}. This means in $\bar{G_o}[V(C)] $ (where $\bar{G_o}[V(C)] $ denotes the subgraph of $ \bar{G_o} $ induced by vertices $ V(C)$, and $\bar{G_o}$ is the complement of the graph $ G_o $ on the same vertices of $ G_o $ such that two distinct vertices of $\bar{G_o} $ are adjacent if and only if they are not adjacent in $ G_o $), there exists a path between any pair of vertices in $ V(C)$. Ong et al.~\cite{lawrencemultisender} demonstrated that we can exploit message-connected cycles for uniprior multi-sender index coding. Now in TSUIC, we show that message-connected cycles can also be exploited in order to get savings\footref{f4} in the following Lemma. \vskip+2pt
\begin{Lemma} \label{lemma1}
	If $ D $ is any message-connected cycle $ C $ with $ N $ vertices, then there exists an index code of length $ N-1 $ satisfying the constraint due to the two senders in $ (D,G_o)$.
\end{Lemma}
\begin{IEEEproof}
	By the definition of a message-connected cycle $ C $, vertices in $ V(C)$ forms a connected component\footnote{If any pair of vertices in a (component of) graph is connected by a path, then the (component of) graph is connected (component of) graph.} in $ \bar{G}_o$ denoted by $ G_c $. One can prove that a graph is connected if and only if it has a spanning tree.\footnote{A spanning tree of a (component of) graph is a tree which includes all of the vertices of the (component of) graph.} So, $ G_c $ has a spanning tree, denoted $ T_c $, which has $ N-1 $ edges and $N$ vertices of $ C$\cite{lawrencemultisender}. Let $ V(C)=\{1,2,\dotsc,N\} $. Now we present a construction of an index code for $ C $ in the following way: $ \{x_i\oplus x_j: (i,j)\in E(T_c) \}$. The index codelength is $ N-1 $ because the number of edges in $ T_c $ is $ N-1 $. For any $ i,j\in V(C) $, if $ (i,j)\in E(T_c) $, then $ (i,j)\in E(G_c)$. Consequently, $ x_i$ and $ x_j $ are located at the same sender. Thus the index code satisfies the constraint due to the two senders. One can verify that any $i\in V(C)$ can decode $ x_i $ from $ \{x_i\oplus x_j: (i,j)\in E(T_c)\}$ and $ \mathcal{H}_i $.
\end{IEEEproof}

Now we prove that if any cycle has at least a vertex requesting a common message, then it is message connected.\vskip+2pt
\begin{Lemma}\label{prop3}
	For a cycle $ C $, if there exists a vertex $ i\in V(C) $ such that $ x_i\in \mathcal{P}_c $, i.e., the vertex is requesting a common message, then $ C $ is message connected in $ (D, G_o) $. 
\end{Lemma}
\begin{IEEEproof}
	If any $ i\in V(C)$ such that $ x_i\in \mathcal{P}_c $, then in $ \bar{G_o} $, $ i $ has an edge with each of the vertices in $ V(C)\setminus\{i\} $. Thus in $ \bar{G_o} $, the graph component formed by vertices in $ V(C)\subseteq V(\bar{G_o}) $ always has a spanning tree. Consequently, the cycle $ C $ is message connected in $ (D, G_o) $.
\end{IEEEproof}

In the following proposition, we exhibit that any non-message-connected cycle is of no use for TSUIC. \vskip+2pt
\begin{proposition} \label{prop4}
	If $ D $ is a cycle $ C $ which is not message connected in $ (D,G_o) $, then no scheme can provide savings\footref{f4} for $ D $ in TSUIC.
\end{proposition}
\begin{IEEEproof}
	For the given $ D $, we have the following: (i) $ \mathcal{P}_c=\emptyset $ because there is no vertex $i\in V(D) $ such that $ x_i\in \mathcal{P}_c $, otherwise $ D $ will be a message-connected cycle from Lemma~\ref{prop3} (contradiction of the given condition), and (ii) $\mathcal{P}_1, \mathcal{P}_2 \neq \emptyset$ because if $ \mathcal{P}_1=\emptyset $, then $\{ x_i:i\in V(D) \}=\mathcal{P}_2$ since $ \mathcal{P}_c=\emptyset $, so $ D $ will be a message-connected cycle (contradiction of the given condition), and similar result holds if $ \mathcal{P}_2=\emptyset$. Thus for a vertex $ i\in V(D) $, $ x_i\in \mathcal{P}_1\cup \mathcal{P}_2$. Now invoking Theorem~\ref{theorem2}, we get $ \beta_t(D,G_o)=\beta_t(D_1)+\beta_t(D_2)$, where $ D_1 $ and $ D_2 $ are the sub-digraphs of $ D $ induced by $ V(D_1)=\{i:x_i\in \mathcal{P}_1 \} $ and $ V(D_2)=\{i:x_i\in \mathcal{P}_2 \} $ respectively. Moreover, $ V(D_1)\cup V(D_2)=V(D) $ because $ \mathcal{P}_c=\emptyset $. The sub-digraphs $ D_1 $ and $ D_2 $ are acyclic because they are strict sub-digraphs of $ C $. Now one can show that $ \mathsf{MAIS}(D_1)=|V(D_1)|$ and $ \mathsf{MAIS}(D_2)=|V(D_2)|$. It has been shown that in SSUIC, $ \mathsf{MAIS}(D)\leq\beta_t(D)$~\cite{maisbound} and $ \beta_t(D)\leq |V(D)| $, so for the sub-digraphs, $ \mathsf{MAIS}(D_1)=|V(D_1)|=\beta_t(D_1)$ and $ \mathsf{MAIS}(D_2)=|V(D_2)|=\beta_t(D_2)$. Altogether, $\beta_t(D,G_o)=\beta_t(D_1)+\beta_t(D_2)=|V(D_1)|+|V(D_2)|=|V(D)|=|V(C)|$.
\end{IEEEproof}

\begin{definition} [Two-sender cycle-covering number, $\ell_{\mathsf{CY}}(D,G_o) $]
	The difference between $ |V(D)| $ and the maximum number of disjoint message-connected cycles in $ (D,G_o) $ is the two-sender cycle-cover number.
\end{definition}

\begin{definition} [Two-sender cycle-cover scheme]
	The two-sender cycle-cover scheme finds a set of disjoint message-connected cycles that contribute to achieve an index code of length $\ell_{\mathsf{CY}}(D,G_o) $, and constructs the index code that has (i) the following coded symbols for each disjoint message-connected cycle $ C $: $ \{x_i\oplus x_j: (i,j)\in E(T_c) \}$, where $ T_c $ is a spanning tree connecting all vertices of $ C $ in $ \bar{G_o}[V(C)]$ (see the proof of Lemma~\ref{lemma1}), and (ii) uncoded messages which are requested by vertices not included in any $ C $ in $ (D,G_o)$. \vskip+2pt
\end{definition}

\begin{theorem} \label{theorem3}
	The optimal broadcast rate of a TSUIC problem is upper bounded by the two-sender cycle-cover number, i.e, $ \beta(D,G_o) \leq \beta_t(D,G_o)\leq  \ell_{\mathsf{CY}}(D,G_o),\ \forall t$.
\end{theorem} 
\begin{IEEEproof}
	Let there be $M$ disjoint message-connected cycles that contribute to achieve an index code of length $\ell_{\mathsf{CY}}(D,G_o) $, where each message-connected cycle $ C_i $ has $ m_i $ vertices (where $ m_i\geq 2 $). In each $ C_i $, by the definition of the two-sender cycle-cover scheme, it constructs $ m_i-1 $ coded symbols. Now $ \ell_{\mathsf{CY}}(D,G_o)=|V(D)|-\sum_{i=1}^{M} m_i+\sum_{i=1}^{M}(m_i-1)=|V(D)|-M$, is an upper bound to the optimal broadcast rate of the TSUIC problem for any $ t $.
\end{IEEEproof}

\vskip+2pt
\begin{corollary}\label{cor1}
	If $ D $ is $ M $ disjoint cycles $ \{C_k: k\in \{1,2,\dotsc, M\}\} $, and each $C_k $ is message connected in $ (D,G_o) $, then $\beta(D,G_o)=\beta_t(D,G_o)=\ell_{\mathsf{CY}}(D,G_o)=|V(D)|-M =\beta(D)=\ell_{\mathsf{CY}}(D) $. 
\end{corollary}
\begin{IEEEproof}
	In SSUIC, for the given $ D $, the single-sender cycle-cover achieves $ \ell_{\mathsf{CY}}(D)=|V(D)|-M$~\cite{chaudhary}. Furthermore, we require to remove at least $ M $ vertices, one from each cycle to make $ D $ acyclic, thus $\mathsf{MAIS}(D)=|V(D)|-M $. It has been shown that in SSUIC, $ \mathsf{MAIS}(D)\leq\beta(D) $~\cite{maisbound}. Altogether, for $ D $ in SSUIC, we get $ \mathsf{MAIS}(D)=\beta(D)=\ell_{\mathsf{CY}}(D)=|V(D)|-M $. Now in the given TSUIC problem, each $C_k$ is message connected, so it follows from the proof of Theorem~\ref{theorem3} that $ \ell_{\mathsf{CY}}(D,G_o)=|V(D)|-M$. On the whole, we get $ \beta(D,G_o)=  \beta_t(D,G_o)=\ell_{\mathsf{CY}}(D,G_o)=|V(D)|-M=\beta(D)=\ell_{\mathsf{CY}}(D)$ because $\beta(D)\leq \beta(D,G_o)\leq \beta_t(D,G_o)\leq \ell_{\mathsf{CY}}(D,G_o)$ and $ \beta(D)= |V(D)|-M=\ell_{\mathsf{CY}}(D,G_o)$.
\end{IEEEproof}

\subsection{Clique cover, its relation with the chromatic number, and its two-sender extension} \label{sectionB}

In SSUIC, the clique-cover scheme~\cite{ISCOD} exploits any cliques in $ D $ to obtain some savings.\footref{f4} Moreover, if $ \rho  $ is a clique with $ V(\rho)=\{1,2\dotsc,n\} $, then every $i\in V(\rho) $ can decode $ x_i $ from one coded symbol $ \bigoplus_{i=1}^n x_i $. This provides the savings\footref{f4} of $ n-1 $ in each $ \rho $. However, in TSUIC, because of the introduction of the constraint due to the two senders, we can only exploit a clique $ \rho $ satisfying the following property: for all $ i,j \in V(\rho)$, there exists no $ (i,j)\in E(G_o) $. In other words, a sender must have messages requested by all of the vertices in the clique. Consequently, the coded symbol $ \bigoplus_{i=1}^n x_i $, constructed for the clique $ \rho $ satisfies the constraint due to the two senders. We call the clique $ \rho $ a \emph{two-sender clique}. 

\begin{definition} [Two-sender clique-cover number, $ \ell_{\mathsf{CL}}(D,G_o) $]
	The two-sender clique-cover number is the minimum number of disjoint two-sender cliques in $ (D,G_o) $. A single vertex is a two-sender clique of size one.
\end{definition}

\begin{definition}[Chromatic number, $ \goodchi(D) $]
The minimum number of colors over all possible proper colorings\footnote{In a proper vertex coloring of a graph, no two vertices connected by one edge share the same color.} of a graph $ D $ is called the chromatic number of the graph. 	
\end{definition}
\vskip+2pt
\subsubsection{SSUIC and the chromatic number}
In SSUIC, it has been shown that for a digraph $D$, the clique-cover scheme achieves the chromatic number of the underlying undirected graph\footnote{The underlying undirected graph of a digraph is the graph constructed by replacing each arc of the digraph by one corresponding edge (undirected).} of $ \bar{D} $, denoted as $ U_{\bar{D}} $. This means, $ \ell_{\mathsf{CL}}(D)=\goodchi(U_{\bar{D}})$~\cite{ISCOD}. 

\vskip+2pt
\subsubsection{TSUIC and the chromatic number}
In TSUIC, we propose the following:
Firstly, we introduce $ G_o $ in $ U_{\bar{D}}$, i.e., we construct $U_{\bar{D},G_o}\triangleq U_{\bar{D}}\cup G_o $. Secondly, we subject $ U_{\bar{D},G_o} $ to a proper coloring. Coloring $ U_{\bar{D},G_o} $ ensures that any two vertices each requesting a private message from a distinct sender will get different colors because there is an edge connecting the corresponding vertices in $ G_o $. Finally, we compute the chromatic number of $ U_{\bar{D},G_o} $. This is called the two-sender chromatic number denoted by $ \goodchi(\bar{D},G_o)$. In a similar way to its SSUIC counterpart, one can verify that 
\begin{equation} \label{eq3}
\ell_{\mathsf{CL}}(D,G_o)=\goodchi(\bar{D},G_o)=\goodchi(U_{\bar{D},G_o}).
\end{equation}

\begin{definition} [Two-sender clique-cover scheme]
	The two-sender clique-cover scheme finds a set of disjoint two-sender cliques that contribute to achieve an index code of length $ \ell_{\mathsf{CL}}(D,G_o) $, and constructs an index code in which the coded symbol for each of the disjoint two-sender cliques is the bit-wise XOR of messages requested by all of the vertices in it. \vskip+2pt
\end{definition}

\begin{theorem} \label{theorem4}
	The optimal broadcast rate of a TSUIC problem is upper bounded by the two-sender clique-covering number, i.e, $ \beta(D,G_o) \leq \beta_t(D,G_o)\leq  \ell_{\mathsf{CL}}(D,G_o)=\goodchi(U_{\bar{D},G_o}),\ \forall t$.
\end{theorem} 
\begin{IEEEproof}
	Let there be $ M $ two-sender cliques that contribute to achieve an index code of length $ \ell_{\mathsf{CL}}(D,G_o) $. The two-sender clique-cover scheme constructs one coded symbol for each of the cliques, so $ \ell_{\mathsf{CL}}(D,G_o)=\sum_{i=1}^{M}1=M$ and this is an upper bound to the optimal broadcast rate of the TSUIC problem for any $ t $. Considering \eqref{eq3}, $\goodchi(U_{\bar{D},G_o})=\ell_{\mathsf{CL}}(D,G_o)=M $.
\end{IEEEproof}
\vskip+2pt
\begin{corollary}\label{cor}
	If $ D $ is $ M $ disjoint cliques $ \{\rho_k:k\in\{1,2,\dotsc,M\}\} $, and each $\rho_k $ is a two-sender clique in $ (D,G_o) $, then $\beta(D,G_o)=\beta_t(D,G_o)=\ell_{\mathsf{CL}}(D,G_o)=M=\beta(D)=\ell_{\mathsf{CL}}(D)$.
\end{corollary}
\begin{IEEEproof}
	In SSUIC, for the given D, the single-sender clique-cover achieves $ \ell_{\mathsf{CL}}(D)=M  $~\cite{ISCOD}. Furthermore, one can show that $ \mathsf{MAIS}(D)=M $. It has been shown that $ \mathsf{MAIS}(D)\leq \beta(D) $~\cite{maisbound}. Altogether, for $ D $ in SSUIC, $   \mathsf{MAIS}(D)=\beta(D)=\ell_{\mathsf{CL}}(D)=M$. Now in the given TSUIC problem, the messages requested by all vertices in each $\rho_k $ are at one sender, so it follows from the proof of Theorem~\ref{theorem4} that $ \ell_{\mathsf{CL}}(D,G_o)=M $. On the whole, we get $\beta(D,G_o)=\beta_t(D,G_o)=\ell_{\mathsf{CL}}(D,G_o)=M=\beta(D)=\ell_{\mathsf{CL}}(D)$ because $ \beta(D)\leq \beta(D,G_o)\leq \beta_t(D,G_o)\leq \ell_{\mathsf{CL}}(D,G_o)$ and $ \beta(D)=M=\ell_{\mathsf{CL}}(D,G_o)$. 
\end{IEEEproof}

\subsection{Local chromatic number and its two-sender extension}

 \begin{figure}[t] 
 	\centering
 	\subfloat []{ %
 		\includegraphics[height=2.9cm,keepaspectratio]{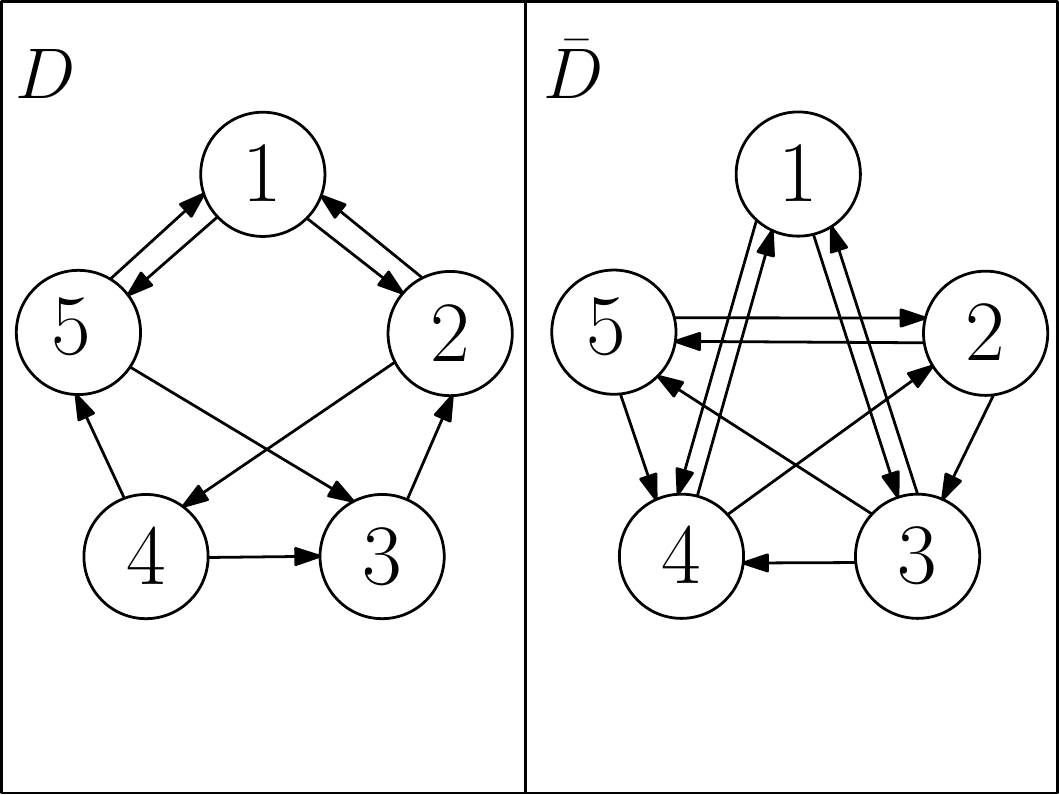}
 		\label{graph1}
 	}
 	\hskip-4pt
  	\subfloat []{
 		\centering
 		\includegraphics[height=2.9cm,keepaspectratio]{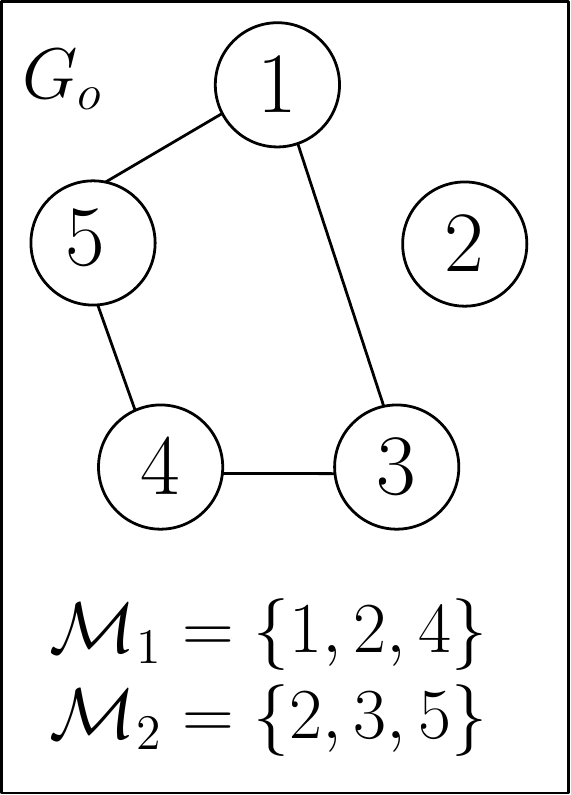}
 		\label{graph2}
 	}
 	\hskip-4pt
  	\subfloat []{
 		\centering
 		\includegraphics[height=2.9cm,keepaspectratio]{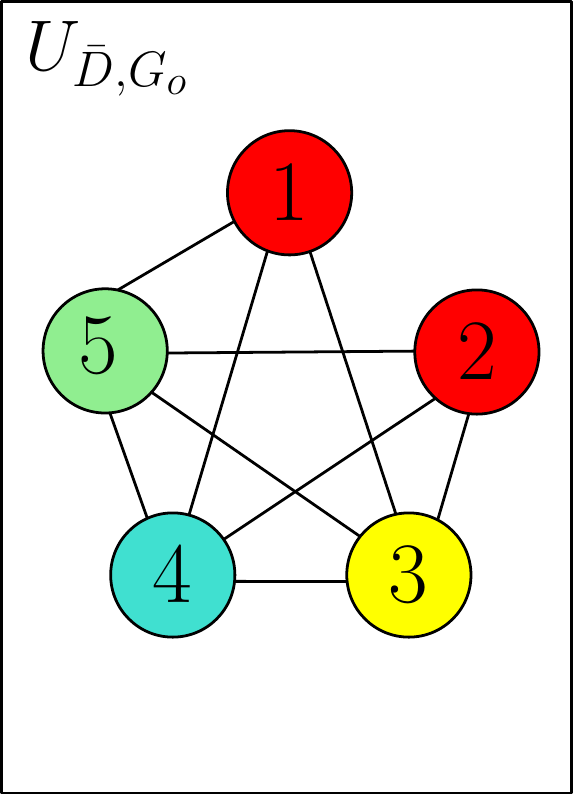}
 		\label{graph3}
 	}
 	\caption{(a) Digraph $ D $ represents an example of an index-coding problem, and $ \bar{D} $ is its complemented digraph, (b) the sender-constraint graph $ G_o $ due to message sets $ \mathcal{M}_1=\{1,2,4\} $ and $ \mathcal{M}_2=\{2,3,5\}$ at the two senders, and (c) a proper coloring of $U_{\bar{D},G_o}=\bar{D}\cup G_o$.} 
 	\label{graphs}
 	\end{figure}

\subsubsection{SSUIC and the local-chromatic number}

The local-chromatic number of the complemented digraph\footref{f3} of a digraph $ D $, denoted by $ \goodchi_{\ell} (\bar{D}) $, is defined as follows: Let $ \mathit{J}:V(\bar{D})\rightarrow  \mathcal{J}$ be a proper coloring of $ \bar{D}$ ignoring the orientation of the arcs (i.e., in $ U_{\bar{D}} $) with a set of colors $ \mathcal{J} $. The local-chromatic number $ \goodchi_{\ell} (\bar{D}):=\underset{\mathit{J}}{\text{min}} \underset{i\in V(\bar{D})}{\text{max}} |\{\mathit{J}(u): u\in N_{\bar{D}}^+ (i)\}|+1 $, where the minimum is taken over all proper colorings of $ U_{\bar{D}}$, and $ N_{\bar{D}}^+ (i) $ is the out-neighborhood of $ i $ in $ \bar{D} $.  
  
 Shanmugam et al.~\cite{localgarphcoloring} proved that for an SSUIC problem represented by $ D $, the optimal broadcast rate is upper bounded by $ \goodchi_{\ell} (\bar{D})$. Moreover, we have $ \goodchi_{\ell} (\bar{D})\leq \goodchi (\bar{D})$. The upper bound $ \goodchi_{\ell} (\bar{D})$ is achieved by the index code generated in the following way: Firstly, consider a $ \goodchi_{\ell} (\bar{D}) \times |\mathcal{J}|$ generator matrix, denoted $ \boldsymbol{G}_{|\mathcal{J}|} $, of a $ (|\mathcal{J}|,\goodchi_{\ell} (\bar{D})) $-MDS code (any $  \goodchi_{\ell} (\bar{D}) $ columns of $ \boldsymbol{G}_{|\mathcal{J}|} $ are linearly independent), over a suitable field $ \mathbb{F}_q $ for a sufficiently large $ q $. Secondly, consider a mapping function, say $ \boldsymbol{b} $, which assigns one column vector of $ \boldsymbol{G}_{|\mathcal{J}|}$ to each color in $ \mathcal{J} $, i.e., $ \boldsymbol{b}:\mathcal{J}\rightarrow \mathcal{C}$, where $\mathcal{C}$ is the set of all of the column vectors of $ \boldsymbol{G}_{|\mathcal{J}|} $. Finally, for $ D $ with $ N $ vertices, we construct a $ \goodchi_{\ell} (\bar{D}) \times N $ matrix $ \boldsymbol{G}_N=[\boldsymbol{b}(\mathit{J}(1))\ \boldsymbol{b}(\mathit{J}(2)) \dotsc \boldsymbol{b}(\mathit{J}(N)) ] $, wherein each column $ i $ corresponds to the vector assigned to the color of the vertex $ i $. If the message alphabet $ \{0,1\}^t $ can be bijectively mapped to the finite field $ \mathbb{F}_q $, then the index code of length $ \goodchi_{\ell} (\bar{D}) $ is constructed by performing $ \boldsymbol{G}_N \bm{\cdot} [x_1\ x_2\ \cdots\ x_N]^\intercal $, where, with an abuse of notation, $ x_i\in \mathbb{F}_q\ \forall x_i\in \mathcal{M} $.  \vskip+2pt       
  \begin{figure*}[t]
  	\centering
  	\hskip-8pt
  	\subfloat []{ %
  		\includegraphics[height=3cm,keepaspectratio]{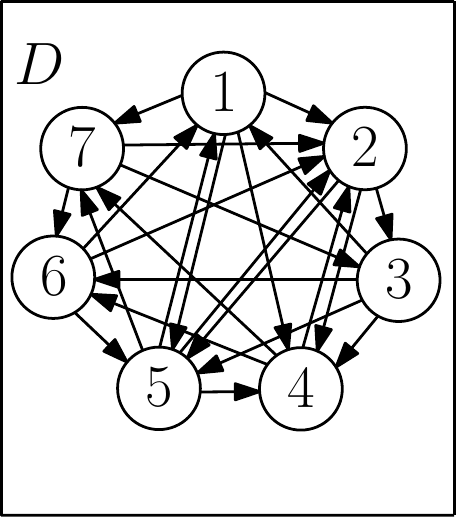}
  		\label{g1}
  	}
  	\hskip-5pt
  	\subfloat []{ %
  		\includegraphics[height=3cm,keepaspectratio]{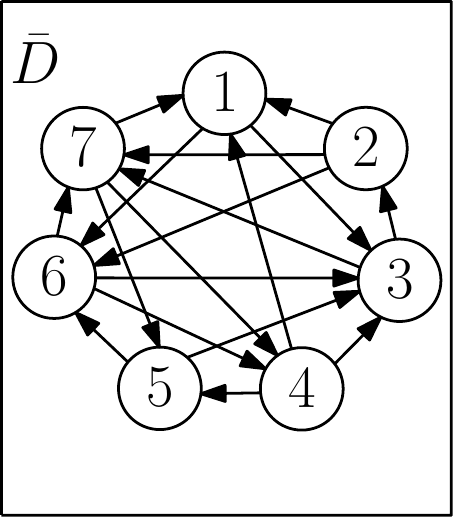}
  		\label{g2}
  	}
  	\hskip-5pt
  	\subfloat []{ %
  		\includegraphics[height=3cm,keepaspectratio]{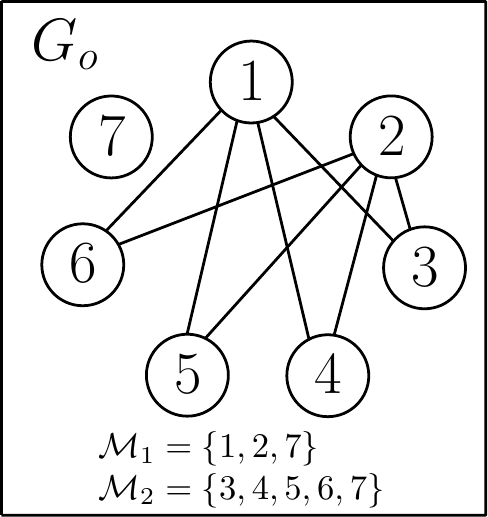}
  		\label{g3}
  	}
  	\hskip-5pt
  	\subfloat []{ %
  		\includegraphics[height=3cm,keepaspectratio]{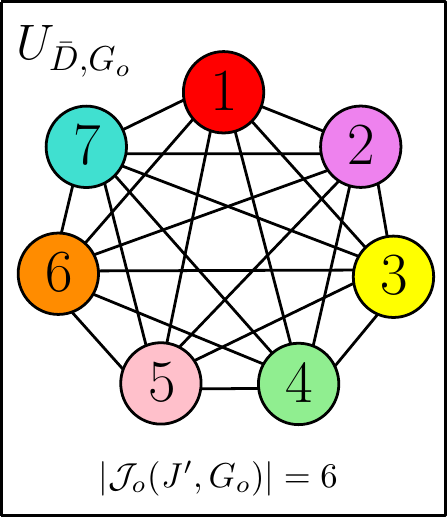}
  		\label{g4}
  	}
  	\hskip-4pt
  	\subfloat []{ %
  		\includegraphics[height=3cm,keepaspectratio]{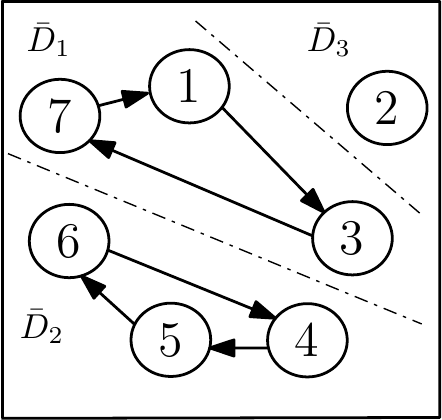}
  		\label{g5}
  	}
  	\hskip-5pt
  	\subfloat []{ %
  		\includegraphics[height=3cm,keepaspectratio]{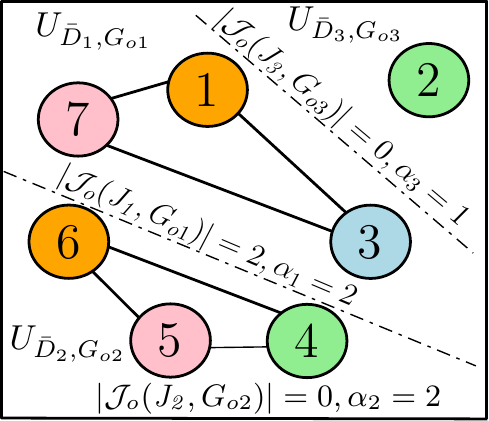}
  		\label{g6}
  	} 	
  	%
  	\caption{(a) The digraph $ D $, (b) the complemented digraph $ \bar{D} $, (c) the sender-constraint graph $ G_o $ for $ \mathcal{M}_1=\{1,2,7\} $ and $ \mathcal{M}_2=\{3,4,5,6,7\} $, (d) a proper coloring of $ U_{\bar{D},G_o} $, (e) $\bar{D} $ is partitioned into three sub-digraphs $ \bar{D}_1,\bar{D}_2\ \text{and}\ \bar{D}_3 $, and (f) the three corresponding properly colored underlying undirected graphs $U_{\bar{D}_1,G_{o1}},\ U_{\bar{D}_2,G_{o2}}\ \text{and}\ U_{\bar{D}_3,G_{o3}} $. For this problem, we have $ \alpha=|\mathcal{J}_o(\mathit{J^*},G_o)|=6 $, whereas after partitioning $ \bar{D} $, we get $ \ell_{p}(\bar{D},G_o)=\alpha_1+\alpha_2+\alpha_3=2+2+1=5$. One can show that $ \ell_{p}(\bar{D},G_o)=5$ is achieved by the index code $ \{ x_1\oplus x_7,\ x_3\oplus x_7,\ x_2,\ x_4\oplus x_6,\ x_5\oplus x_6\} $.}  
  	\label{gs}
  \end{figure*}
 \subsubsection{TSUIC and the local-chromatic number}
 In TSUIC, we propose the following: Firstly, we properly color $ U_{\bar{D},G_o} $ in the same way as the coloring done for finding the chromatic number in a TSUIC problem (refer to Section~\ref{sectionB}). Secondly, noting that $V(\bar{D})=V(U_{\bar{D},G_o})$, let $ \mathit{J}':V(\bar{D})\rightarrow  \mathcal{J'}$ be a proper coloring of $ U_{\bar{D},G_o} $ with a set of colors $ \mathcal{J'} $ (for an illustration, refer to Fig.~\ref{graphs}). Finally, we define the two-sender local-chromatic number, $ \goodchi_{\ell} (\bar{D}, G_o):=\underset{\mathit{J}'}{\text{min}} \underset{i\in V(\bar{D})}{\text{max}} |\{\mathit{J}'(u): u\in N_{\bar{D}}^+ (i)\}|+1 $, where the minimum is taken over all proper colorings of $ U_{\bar{D},G_o}$ (for example, the problem in  Fig.~\ref{graphs} has $ \goodchi_{\ell} (\bar{D}, G_o)=4$). In TSUIC, we attempt to construct an index code in the similar way to SSUIC. However, because of the constraint due to the two senders, the mapping function that maps each color to one column vector from a suitable MDS-code generator matrix should satisfy the following property: If the colors of any two vertices\footnote{Any two vertices connected by an edge in $ G_o $ have two different colors in the properly colored $ U_{\bar{D},G_o}$.} connected by an edge in $ G_o $ are mapped to vectors $ \boldsymbol{a}=[a_1\ a_2\ \cdots a_n] $ and $ \boldsymbol{a'}=[a'_1\ a'_2\ \cdots a'_n] $ respectively for some integer $ n $, then for $ i\in \{1,2,\dotsc,n\} $, if any element $ a_i\neq 0 $, then the respective $ a'_i=0 $, and vice versa. There can exist many cases where the mapping function would not satisfy this property for $ n=\goodchi_{\ell} (\bar{D}, G_o) $. Thus the length $\goodchi_{\ell} (\bar{D}, G_o) $ is not always achievable in $ (\bar{D}, G_o) $. Now we present an achievable index codelength for TSUIC by modifying the single-sender local-chromatic approach.
 
 At first, we define the following: (i) For any coloring scheme $ \mathit{J'} $, let the set of colors of all of the vertices having at least an edge connecting them in $ G_o $ be $  \mathcal{J}_o(\mathit{J'},G_o)$, i.e., $\mathcal{J}_o(\mathit{J'},G_o)\triangleq \{\mathit{J}'(i): (i,j)\in E(G_o)\}$ (clearly, $  \mathcal{J}_o(\mathit{J'},G_o)\subseteq \mathcal{J'}$), (ii) $N_{\ell}(\mathit{J'},\bar{D},G_o)\triangleq \underset{i\in V(\bar{D})}{\text{max}} |\{\mathit{J}'(u): u\in N_{\bar{D}}^+ (i)\}|+1$, and (iii) for $ (D,G_o) $, let a coloring scheme that minimizes the maximum of $ N_{\ell}(\mathit{J'},\bar{D},G_o)$ and $|\mathcal{J}_o(\mathit{J'},G_o)|$ over all proper colorings of $ U_{\bar{D},G_o}$ be denoted by $ J^*$ and $J^*:V(\bar{D})\rightarrow \mathcal{J}^*$.  

\vskip+2pt
\begin{theorem}\label{th6}
	The optimal broadcast rate $\beta({D},G_o)\leq \text{max} \left\{N_{\ell}(\mathit{J^*},\bar{D},G_o),\ |\mathcal{J}_o(\mathit{J^*},G_o)| \right\} \leq \goodchi(\bar{D},G_o)$.  
\end{theorem}
\begin{IEEEproof}
	For a proper coloring of a digraph, the local-chromatic number is always less than or equal to the chromatic number~\cite{localgarphcoloring}. Likewise, in $ (D,G_o) $, we have $\goodchi_{\ell}(\bar{D},G_o)\leq \goodchi(\bar{D},G_o)$. Thus $ N_{\ell}(\mathit{J^*},\bar{D},G_o) \leq \goodchi(\bar{D},G_o)=|\mathcal{J}^*|$. By the definition of $\mathcal{J}_o(\mathit{J'},G_o) $, we have $|\mathcal{J}_o(\mathit{J^*},G_o)|\leq |\mathcal{J}^*|$. Altogether, we get $ \text{max} \left\{N_{\ell}(\mathit{J^*},\bar{D},G_o),\ |\mathcal{J}_o(\mathit{J^*},G_o)| \right\} \leq \goodchi(\bar{D},G_o) $.

	Now we prove the first inequality; let $ \alpha= \text{max} \left\{N_{\ell}(\mathit{J^*},\bar{D},G_o),\ |\mathcal{J}_o(\mathit{J^*},G_o)| \right\}$, we will show that there exist an index code of length $\alpha $. The basic steps are similar to the proof presented by Shanmugam et al.~\cite[Theorem 1]{localgarphcoloring} except the proof of the satisfaction of the constraint due to the two senders. Let $ \boldsymbol{G}_{|\mathcal{J}^*|} $ be a $ \alpha \times |\mathcal{J}^*| $ generator matrix of a $ (|\mathcal{J}^*|,\alpha)$-MDS code over a suitable field $ \mathbb{F}_q $ for a sufficiently large $ q $. Now writing $  \boldsymbol{G}_{|\mathcal{J}^*|} $ in its standard form,\footnote{Any generator matrix of an MDS code can be written in its standard form (or systematic form).} we get $ \boldsymbol{G}_{|\mathcal{J}^*|}=[\boldsymbol{I}_{\alpha}|\boldsymbol{P}]$, where $ \boldsymbol{I}_{\alpha} $ is an $ \alpha \times \alpha $ identity matrix, and $ \boldsymbol{P} $ is an $ \alpha \times (|\mathcal{J}^*|-\alpha ) $ matrix that has non-zero elements in all of its coordinates because every square sub-matrix of $ \boldsymbol{P} $ is non-singular for the generator matrix of an MDS code. Without loss of generality, let $ \mathcal{J}_o(\mathit{J}^*,G_o)=\{1,2,\dotsc,|\mathcal{J}_o(\mathit{J^*},G_o)|\}$, and $ \boldsymbol{c}'_i $ be the $ i $-th column vector of $ \boldsymbol{G}_{|\mathcal{J}^*|}=[\boldsymbol{I}_{\alpha}|\boldsymbol{P}]$. Moreover, the column vector $ \boldsymbol{c}'_i$ for $ i\in \{1,2,\dotsc,\alpha\} $ is an $ i $-th normal basis vector.  
	We now define a mapping function which assigns each color in $ \mathcal{J}^* $ to one unique column vector of $ \boldsymbol{G}_{|\mathcal{J}^*|}=[\boldsymbol{I}_{\alpha}|\boldsymbol{P}]$. Let that function be $ \boldsymbol{b}':\mathcal{J}^*\rightarrow \mathcal{C}'$, where $\mathcal{C}'$ is the set of column vectors of $ \boldsymbol{G}_{|\mathcal{J}^*|}=[\boldsymbol{I}_{\alpha}|\boldsymbol{P}]$, such that we map first $ |\mathcal{J}_o(\mathit{J^*},G_o)|$ columns of $ \boldsymbol{G}_{|\mathcal{J}^*|}=[\boldsymbol{I}_{\alpha}|\boldsymbol{P}]$ to each color in $ \mathcal{J}_o(\mathit{J^*},G_o)$, i.e., $ \boldsymbol{b}'(\mathit{J}^*(i))= \boldsymbol{c}'_i$ for $ i\in \{1,2,\dotsc,|\mathcal{J}_o(\mathit{J^*},G_o)|\} $, and the rest of $|\mathcal{J^*}|-|\mathcal{J}_o(\mathit{J^*},G_o)|$ colors in $\mathcal{J}^*$ are arbitrarily mapped to the remaining $|\mathcal{J}^*|-|\mathcal{J}_o(\mathit{J^*},G_o)|$ column vectors. Now for $ D $ with $ N$ vertices, we construct an $ \alpha \times N $ matrix $ \boldsymbol{G}_N=[\boldsymbol{b'}(\mathit{J^*}(1))\ \boldsymbol{b}'(\mathit{J^*}(2))\ \cdots\ \boldsymbol{b}'(\mathit{J^*}(N))] $, where each column $ i $ corresponds to the vector assigned to the color of vertex $ i $. If the message alphabet $ \{0,1\}^t $ can be bijectively mapped to the finite field $ \mathbb{F}_q $, then the index code of length $\alpha $ is constructed by performing $ \boldsymbol{G}_N \bm{\cdot} [x_1\ x_2\ \cdots\ x_N]^\intercal $, and it satisfies the constraint due to the two senders because while mapping colors to vectors, we mapped each color in $\mathcal{J}_o(\mathit{J^*},G_o)$ to one unique orthonormal basis vector of $ \boldsymbol{G}_{|\mathcal{J}^*|}=[\boldsymbol{I}_{\alpha}|\boldsymbol{P}]$.
 
 For any $ i\in V(\bar{D}) $ and $ k\in N^+_{\bar{D}}(i)\setminus \{i\} $, $ \boldsymbol{b'}(\mathit{J^*}(i)) $ is different from ${ \boldsymbol{b'}(\mathit{J^*}(k)) }$. This is because $ \mathit{J^*}$ represents a proper coloring of $ U_{\bar{D},G_o} $. From the MDS property, any $\alpha$ columns of the $\boldsymbol{G}_{|\mathcal{J}^*|}=[\boldsymbol{I}_{\alpha}|\boldsymbol{P}] $ are linearly independent, and for any vertex $ i $, the number of colors in any closed out-neighborhood is at most $ N_{\ell}(\mathit{J^*},\bar{D},G_o)$, i.e., $| \mathit{J^*}(N^+_{\bar{D}}(i))|\leq N_{\ell}(\mathit{J^*},\bar{D},G_o) $. Thus $ \boldsymbol{b'}(\mathit{J^*}(i)) $ is linearly independent from $ \{\boldsymbol{b'}(\mathit{J^*}(k)) \}_{k\in N^+_{\bar{D}}(i)\setminus \{i\}} $. Consequently, any receiver $ i $ would be able to decode its requested messages from the index code of length $\alpha$ generated by $ {\boldsymbol{G}_N \bm{\cdot} [x_1\ x_2\ \cdots\ x_N]^\intercal}$ and $\mathcal{H}_i$.       
 \end{IEEEproof}

 Now we show that the partitioned approach can help to acquire a better upper bound in general.
 \vskip+2pt
 
\subsubsection{Partitioned local-chromatic approach to TSUIC}
 	For some positive integer $ m $, let $ \bar{D}_1,\bar{D}_2,\dotsc, \bar{D}_m $ be the sub-graphs partitioning $ \bar{D} $ in such a way that $ {\bigcup}_{\forall i \in \{1,2,\dotsc,m\}} V(\bar{D}_i)=V(\bar{D}) $ and $ V(\bar{D}_i)~\cap~V(\bar{D}_j)=~\emptyset$, $\forall i,j\in \{1,2,\dotsc,m \} $ and $ i\neq j $. Let $ G_{o1},G_{o2},\dotsc,G_{om} $ be the sender-constraint graphs of the respective sub-graphs partitioning $ \bar{D} $. Now for $ i\in \{1,2,\cdots,m\} $, we define $ \ell[\{\bar{D}_i,G_{oi}\}_{\forall i}]\triangleq \sum_{i=1}^{m} \alpha_i $, 	
 	where $ \alpha_i=\underset{\mathit{J}_i}{\text{min}}\ \text{max} \left\{N_{\ell}(\mathit{J}_i,\bar{D}_i,G_{oi}),\ |\mathcal{J}_o(\mathit{J}_i,G_{oi})| \right\}$, and it is achievable in each $ D_i $ based on Theorem~\ref{th6}. Now the codelength after partitioning is given by
 	\begin{equation} \label{eqpartitionlc2}
 	\ell_{p}(\bar{D},G_o)= \underset{\{ \bar{D}_i,G_{oi}\}_{\forall i}} {\text{min}}\ \ell[\{\bar{D}_i,G_{oi}\}_{\forall i}], 
 	\end{equation}
 	where the minimum is taken over all partitions of $\bar{D}$.
 	
  \begin{theorem}\label{th5}
  	The optimal broadcast rate $\beta({D},G_o)\leq \ell_{p}(\bar{D},G_o)\leq \alpha=\text{max} \left\{N_{\ell}(\mathit{J^*},\bar{D},G_o),\ |\mathcal{J}_o(\mathit{J^*},G_o)| \right\}$.
  \end{theorem}
  \begin{IEEEproof}
  	The proof follows from the Theorem~\ref{th6} and the definition of $  \ell_{p}(\bar{D},G_o) $, which is obtained by minimizing the index codelength over all partitions of $ \bar{D} $ and $ G_o $ (including the non-partitioned $ \bar{D} $ and $ G_o $ among all partitions).
  \end{IEEEproof}
  We illustrate Theorem~\ref{th5} with one example depicted in Fig.~\ref{gs}. In the example, $ \ell_{p}(\bar{D},G_o)$ is strictly less than $ \alpha$. Note that this way of partitioning is not possible with the trivial extension of the local-chromatic scheme stated in Proposition~\ref{prop2}.
\bibliographystyle{IEEEtran}
\balance





\end{document}